\documentclass[aps,twocolumn,prd,nofootinbib]{revtex4}
\usepackage{amsmath}
\usepackage{graphicx}
\usepackage{dcolumn}
\usepackage{bm}
\usepackage{amssymb}
\usepackage{latexsym}

\newcommand{\be}{\begin{equation}}
\newcommand{\ee}{\end{equation}}
\newcommand{\bq}{\begin{eqnarray}}
\newcommand{\eq}{\end{eqnarray}}

\begin{document}

\title{Reconstructing holographic quintessence}

\author{Xin Zhang}

\affiliation{Institute of
Theoretical Physics, Chinese Academy of Sciences, P.O.Box 2735,
Beijing 100080, People's Republic of China}
\affiliation{Interdisciplinary Center of Theoretical Studies,
Chinese Academy of Sciences, P.O.Box 2735, Beijing 100080, People's
Republic of China}

\begin{abstract}
The holographic dark energy model is an attempt for probing the
nature of dark energy within the framework of quantum gravity. The
dimensionless parameter $c$ determines the main property of the
holographic dark energy. With the choice of $c\geqslant 1$, the
holographic dark energy can be described completely by a
quintessence scalar field. In this paper, we show this
quintessential description of the holographic dark energy with
$c\geqslant 1$ and reconstruct the potential of the quintessence as
well as the dynamics of the scalar field.

\end{abstract}

\maketitle


It has been confirmed admittedly that our universe is experiencing
an accelerating expansion at the present time, by many cosmological
experiments, such as observations of large scale structure (LSS)
\cite{LSS}, searches for type Ia supernovae (SNIa) \cite{SN}, and
measurements of the cosmic microwave background (CMB) anisotropy
\cite{CMB}. This cosmic acceleration observed strongly supports the
existence of a mysterious exotic matter, dark energy, with large
enough negative pressure, whose energy density has been a dominative
power of the universe. The astrophysical feature of dark energy is
that it remains unclustered at all scales where gravitational
clustering of baryons and nonbaryonic cold dark matter can be seen.
Its gravity effect is shown as a repulsive force so as to make the
expansion of the universe accelerate when its energy density becomes
dominative power of the universe. The combined analysis of
cosmological observations suggests that the universe is spatially
flat, and consists of about $70\%$ dark energy, $30\%$ dust matter
(cold dark matter plus baryons), and negligible radiation. Although
we can affirm that the ultimate fate of the universe is determined
by the feature of dark energy, the nature of dark energy as well as
its cosmological origin remain enigmatic at present. However, we
still can propose some candidates to interpret or describe the
properties of dark energy. The most obvious theoretical candidate of
dark energy is the cosmological constant $\lambda$ (vacuum energy)
\cite{Einstein:1917,cc} which has the equation of state $w=-1$.
However, as is well known, there are two difficulties arise from the
cosmological constant scenario, namely the two famous cosmological
constant problems --- the ``fine-tuning'' problem and the ``cosmic
coincidence'' problem \cite{coincidence}. The fine-tuning problem
asks why the vacuum energy density today is so small compared to
typical particle scales. The vacuum energy density is of order
$10^{-47} {\rm GeV}^4$, which appears to require the introduction of
a new mass scale 14 or so orders of magnitude smaller than the
electroweak scale. The second difficulty, the cosmic coincidence
problem, says: Since the energy densities of vacuum energy and dark
matter scale so differently during the expansion history of the
universe, why are they nearly equal today? To get this coincidence,
it appears that their ratio must be set to a specific, infinitesimal
value in the very early universe.

Theorists have made lots of efforts to try to resolve the
cosmological constant problem, but all these efforts were turned out
to be unsuccessful. Of course the theoretical consideration is still
in process and has made some progresses. In recent years, many
string theorists have devoted to understand and shed light on the
cosmological constant or dark energy within the string framework.
The famous Kachru-Kallosh-Linde-Trivedi (KKLT) model \cite{kklt} is
a typical example, which tries to construct metastable de Sitter
vacua in the light of type IIB string theory. Furthermore, string
landscape idea \cite{landscape} has been proposed for shedding light
on the cosmological constant problem based upon the anthropic
principle and multiverse speculation. However, there remain other
candidates to explaining dark energy.

An alternative proposal for dark energy is the dynamical dark energy
scenario. The cosmological constant puzzles may be better
interpreted by assuming that the vacuum energy is canceled to
exactly zero by some unknown mechanism and introducing a dark energy
component with a dynamically variable equation of state. The
dynamical dark energy proposal is often realized by some scalar
field mechanism which suggests that the energy form with negative
pressure is provided by a scalar field evolving down a proper
potential. Actually, this mechanism is enlightened to a great extent
by the inflationary cosmology. As we have known, the occurrence of
the current accelerating expansion of the universe is not the first
time in the expansion history of the universe. There is significant
observational evidence strongly supporting that the universe
underwent an early inflationary epoch, over sufficiently small time
scales, during which its expansion rapidly accelerated under the
driven of an ``inflaton'' field which had properties similar to
those of a cosmological constant. The inflaton field, to some
extent, can be viewed as a kind of dynamically evolving dark energy.
Hence, the scalar field models involving a minimally coupled scalar
field are proposed, inspired by inflationary cosmology, to construct
dynamically evolving models of dark energy. The only difference
between the dynamical scalar-field dark energy and the inflaton is
the energy scale they possess. Famous examples of scalar-field dark
energy models include quintessence \cite{quintessence}, $K$-essence
\cite{kessence}, tachyon \cite{tachyon}, phantom \cite{phantom},
ghost condensate \cite{ghost1,ghost2} and quintom \cite{quintom},
and so forth. Generically, there are two points of view on the
scalar-field models of dynamical dark energy. One viewpoint regards
the scalar field as a fundamental field of the nature. The nature of
dark energy is, according to this viewpoint, completely attributed
to some fundamental scalar field which is omnipresent in
supersymmetric field theories and in string/M theory. The other
viewpoint supports that the scalar field model is an effective
description of an underlying theory of dark energy. On the whole, it
seems that the latter is the mainstream point of view. Since we
regard the scalar field model as an effective description of an
underlying theory of dark energy, a question arises asking: What is
the underlying theory of the dark energy? Of course, hitherto, this
question is far beyond our present knowledge, because that we can
not entirely understand the nature of dark energy before a complete
theory of quantum gravity is established. However, although we are
lacking a quantum gravity theory today, we still can make some
attempts to probe the nature of dark energy according to some
principles of quantum gravity. The holographic dark energy model is
just an appropriate example, which is constructed in the light of
the holographic principle of quantum gravity theory. That is to say,
the holographic dark energy model possesses some significant
features of an underlying theory of dark energy.

The distinctive feature of the cosmological constant or vacuum
energy is that its equation of state is always exactly equal to
$-1$. However, when considering the requirement of the holographic
principle originating from the quantum gravity speculation, the
vacuum energy will become dynamically evolving dark energy.
Actually, the dark energy problem may be in principle a problem
belongs to quantum gravity \cite{Witten:2000zk}. In the classical
gravity theory, one can always introduce a cosmological constant to
make the dark energy density be an arbitrary value. However, a
complete theory of quantum gravity should be capable of making the
property of dark energy, such as the energy density and the equation
of state, be determined definitely and uniquely
\cite{Witten:2000zk}. Currently, an interesting attempt for probing
the nature of dark energy within the framework of quantum gravity is
the so-called ``holographic dark energy'' proposal
\cite{Cohen:1998zx,Horava:2000tb,Hsu:2004ri,Li:2004rb}. It is well
known that the holographic principle is an important result of the
recent researches for exploring the quantum gravity (or string
theory) \cite{holoprin}. This principle is enlightened by
investigations of the quantum property of black holes. Roughly
speaking, in a quantum gravity system, the conventional local
quantum field theory will break down. The reason is rather simple:
For a quantum gravity system, the conventional local quantum field
theory contains too many degrees of freedom, and so many degrees of
freedom will lead to the formation of black hole so as to break down
the effectiveness of the quantum field theory.

For an effective field theory in a box of size $L$, with UV cut-off
$\Lambda$ the entropy $S$ scales extensively, $S\sim L^3\Lambda^3$.
However, the peculiar thermodynamics of black hole \cite{bh} has led
Bekenstein to postulate that the maximum entropy in a box of volume
$L^3$ behaves nonextensively, growing only as the area of the box,
i.e. there is a so-called Bekenstein entropy bound, $S\leq
S_{BH}\equiv\pi M_{\rm Pl}^2L^2$. This nonextensive scaling suggests
that quantum field theory breaks down in large volume. To reconcile
this breakdown with the success of local quantum field theory in
describing observed particle phenomenology, Cohen et al.
\cite{Cohen:1998zx} proposed a more restrictive bound -- the energy
bound. They pointed out that in quantum field theory a short
distance (UV) cut-off is related to a long distance (IR) cut-off due
to the limit set by forming a black hole. In other words, if the
quantum zero-point energy density $\rho_{\rm de}$ is relevant to a
UV cut-off, the total energy of the whole system with size $L$
should not exceed the mass of a black hole of the same size, thus we
have $L^3\rho_{\rm de}\leq LM_{\rm Pl}^2$. This means that the
maximum entropy is in order of $S_{BH}^{3/4}$. When we take the
whole universe into account, the vacuum energy related to this
holographic principle \cite{holoprin} is viewed as dark energy,
usually dubbed holographic dark energy. The largest IR cut-off $L$
is chosen by saturating the inequality so that we get the
holographic dark energy density
\begin{equation}
\rho_{\rm de}=3c^2M_{\rm Pl}^2L^{-2}~,\label{de}
\end{equation} where $c$ is a numerical constant, and $M_{\rm Pl}\equiv 1/\sqrt{8\pi
G}$ is the reduced Planck mass. If we take $L$ as the size of the
current universe, for instance the Hubble scale $H^{-1}$, then the
dark energy density will be close to the observational result.
However, Hsu \cite{Hsu:2004ri} pointed out that this yields a wrong
equation of state for dark energy. Li \cite{Li:2004rb} subsequently
proposed that the IR cut-off $L$ should be taken as the size of the
future event horizon
\begin{equation}
R_{\rm eh}(a)=a\int_t^\infty{dt'\over a(t')}=a\int_a^\infty{da'\over
Ha'^2}~.\label{eh}
\end{equation} Then the problem can be solved nicely and the
holographic dark energy model can thus be constructed successfully.
The holographic dark energy scenario may provide simultaneously
natural solutions to both dark energy problems as demonstrated in
Ref.\cite{Li:2004rb}. The holographic dark energy model has been
tested and constrained by various astronomical observations
\cite{obs1,obs2,obs3}. For other extensive studies, see e.g.
\cite{holoext}.

\begin{figure}[htbp]
\begin{center}
\includegraphics[scale=0.90]{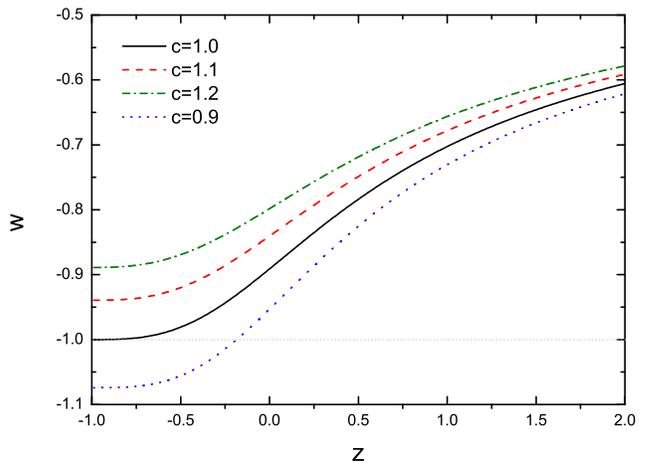}
\caption[]{\small The evolutions of the equation of state of
holographic dark energy. Here we take $\Omega_{\rm m0}=0.3$, and
show the cases for $c=1.0$, 1.1, 1.2 and 0.9. Clearly, the cases in
$c\geq 1$ behave as holographic quintessence, and the case of $c<1$
behaves as holographic quintom.}\label{fig:wevolv}
\end{center}
\end{figure}

Consider now a spatially flat FRW (Friedmann-Robertson-Walker)
universe with matter component $\rho_{\rm m}$ (including both baryon
matter and cold dark matter) and holographic dark energy component
$\rho_{\rm de}$, the Friedmann equation reads
\begin{equation}
3M_{\rm Pl}^2H^2=\rho_{\rm m}+\rho_{\rm de}~,
\end{equation} or equivalently,
\begin{equation}
E(z)\equiv {H(z)\over H_0}=\left(\Omega_{\rm m0}(1+z)^3\over
1-\Omega_{\rm de}\right)^{1/2},\label{Ez}
\end{equation} where $z=(1/a)-1$ is the redshift of the universe.
Note that we always assume spatial flatness throughout this paper as
motivated by inflation. Combining the definition of the holographic
dark energy (\ref{de}) and the definition of the future event
horizon (\ref{eh}), we derive
\begin{equation}
\int_a^\infty{d\ln a'\over Ha'}={c\over Ha\sqrt{\Omega_{\rm
de}}}~.\label{rh}
\end{equation} We notice that the Friedmann
equation (\ref{Ez}) implies
\begin{equation}
{1\over Ha}=\sqrt{a(1-\Omega_{\rm de})}{1\over H_0\sqrt{\Omega_{\rm
m0}}}~.\label{fri}
\end{equation} Substituting (\ref{fri}) into (\ref{rh}), one
obtains the following equation
\begin{equation}
\int_x^\infty e^{x'/2}\sqrt{1-\Omega_{\rm de}}dx'=c
e^{x/2}\sqrt{{1\over\Omega_{\rm de}}-1}~,
\end{equation} where $x=\ln a$. Then taking derivative with respect to $x$ in both
sides of the above relation, we get easily the dynamics satisfied by
the dark energy, i.e. the differential equation about the fractional
density of dark energy,
\begin{equation}
\Omega'_{\rm de}=-(1+z)^{-1}\Omega_{\rm de}(1-\Omega_{\rm
de})\left(1+{2\over c}\sqrt{\Omega_{\rm de}}\right),\label{deq}
\end{equation}
where the prime denotes the derivative with respect to the redshift
$z$. This equation describes behavior of the holographic dark energy
completely, and it can be solved exactly \cite{Li:2004rb}. From the
energy conservation equation of the dark energy, the equation of
state of the dark energy can be given \cite{Li:2004rb}
\begin{equation}
w=-1-{1\over 3}{d\ln\rho_{\rm de}\over d\ln a}=-{1\over 3}(1+{2\over
c}\sqrt{\Omega_{\rm de}})~.\label{w}
\end{equation} Note that the formula
$\rho_{\rm de}={\Omega_{\rm de}\over 1-\Omega_{\rm de}}\rho_{\rm
m}^0a^{-3}$ and the differential equation of $\Omega_{\rm de}$
(\ref{deq}) are used in the second equal sign. It can be seen
clearly that the equation of state of the holographic dark energy
evolves dynamically and satisfies $-(1+2/c)/3\leq w\leq -1/3$ due to
$0\leq\Omega_{\rm de}\leq 1$. Hence, we see clearly that when taking
the holographic principle into account the vacuum energy becomes
dynamically evolving dark energy. The parameter $c$ plays a
significant role in this model. If one takes $c=1$, the behavior of
the holographic dark energy will be more and more like a
cosmological constant with the expansion of the universe, such that
ultimately the universe will enter the de Sitter phase in the far
future. As is shown in \cite{Li:2004rb}, if one puts the parameter
$\Omega_{\rm de}^0=0.73$ into (\ref{w}), then a definite prediction
of this model, $w_0=-0.903$, will be given. On the other hand, if
$c<1$, the holographic dark energy will exhibit appealing behavior
that the equation of state crosses the ``cosmological-constant
boundary'' (or ``phantom divide'') $w=-1$ during the evolution. This
kind of dark energy is referred to as ``quintom'' \cite{quintom}
which is slightly favored by current observations
\cite{cross1,cross2}. For extensive studies on quintom model see
e.g. \cite{quintom2}. If $c>1$, the equation of state of dark energy
will be always larger than $-1$ such that the universe avoids
entering the de Sitter phase and the Big Rip phase. Hence, we see
explicitly, the value of $c$ is very important for the holographic
dark energy model, which determines the feature of the holographic
dark energy as well as the ultimate fate of the universe. As an
illustrative example, we plot in figure \ref{fig:wevolv} the
selected evolutions of the equation of state of holographic dark
energy. We show in the plot the cases of $c=1.0$, 1.1, 1.2 and 0.9.
It is clear to see that the cases in $c\geq 1$ always evolve in the
region of $w\geq -1$, whereas the case of $c<1$ behaves as a quintom
whose equation of state $w$ crosses the cosmological constant
boundary $-1$ during the evolution.

As has been analyzed above, the holographic dark energy scenario
reveals the dynamical nature of the vacuum energy. When taking the
holographic principle into account, the vacuum energy density will
evolve dynamically. On the other hand, as has already mentioned, the
scalar field dark energy models are often viewed as effective
description of the underlying theory of dark energy. However, the
underlying theory of dark energy can not be achieved before a
complete theory of quantum gravity is established. We can,
nevertheless, speculate on the underlying theory of dark energy by
taking some principles of quantum gravity into account. The
holographic dark energy model is no doubt a tentative in this way.
We are now interested in that if we assume the holographic vacuum
energy scenario as the underlying theory of dark energy, how the
scalar field model can be used to effectively describe it.

The quintessence scalar field $\phi$ evolves in its potential
$V(\phi)$ and seeks to roll towards the minimum of the potential,
according to the Klein-Gordon equation
$\ddot{\phi}+3H\dot{\phi}=-dV/d\phi$. The rate of evolution is
driven by the slope of the potential and damped by the cosmic
expansion through the Hubble parameter $H$. The energy density and
pressure are $\rho_\phi=\dot{\phi}^2/2+V$,
$p_\phi=\dot{\phi}^2/2-V$, so that the equation of state of
quintessence $w_\phi=p_\phi/\rho_\phi$ evolves in a region of
$-1<w_\phi<1$. Usually, for making the universe's expansion
accelerate, it should be required that $w_\phi$ must satisfy
$w_\phi<-1/3$. Nevertheless, it can be seen clearly that the
quintessence scalar field can not realize the equation of state
crossing $-1$ \cite{Hu:2004kh}. Therefore, only the holographic dark
energy in cases of $c\geq 1$ can be described by the quintessence.

Now let us see the constraint results for the holographic dark
energy model from the observational data. When combining the
information from SNIa \cite{Riess:2004nr}, CMB \cite{CMB} and LSS
\cite{BAO}, the fitting for the holographic dark energy model gives
the parameter constraints in 1 $\sigma$: $c=0.81^{+0.23}_{-0.16}$,
$\Omega_{\rm m0}=0.28\pm 0.03$, with $\chi_{\rm min}^2=176.67$
\cite{obs1}. In this joint analysis, the SNIa data come from the 157
``gold'' data \cite{Riess:2004nr} including 14 high redshift data
from the {\it Hubble Space Telescope} (HST)/Great Observatories
Origins Deep Survey (GOODS) program and previous data, the CMB
information comes from the measured value of the CMB shift parameter
$R$ given by \cite{CMB} $R\equiv \Omega_{\rm m0}^{1/2}\int_0^{z_{\rm
CMB}}dz'/E(z')=1.716\pm 0.062$, where $z_{\rm CMB}=1089$ is the
redshift of recombination, and the LSS information is provided by
the baryon acoustic oscillation (BAO) measurement \cite{BAO}
$A\equiv \Omega_{\rm m0}^{1/2} E(z_{\rm BAO})^{-1/3}[(1/z_{\rm
BAO})\int_0^{z_{\rm BAO}}dz'/E(z')]^{2/3}=0.469\pm 0.017$, where
$z_{\rm BAO}=0.35$. Furthermore, the X-ray gas mass fraction of rich
clusters, as a function of redshift, has also been used to constrain
the holographic dark energy model \cite{obs2}. The $f_{\rm gas}$
values are provided by {\it Chandra} observational data, the X-ray
gas mass fraction of 26 rich clusters, released by Allen et al.
\cite{Allen:2004cd}. The main results, i.e. the 1 $\sigma$ fit
values for $c$ and $\Omega_{\rm m0}$ are: $c=0.61^{+0.45}_{-0.21}$
and $\Omega_{\rm m0}=0.24^{+0.06}_{-0.05}$, with the best-fit
chi-square $\chi_{\rm min}^2=25.00$ \cite{obs2}. We see that,
basically, in one-sigma error range, the holographic dark energy
will behave as quintom-like dark energy whose equation-of-state
crosses the $w=-1$ line during the evolution.\footnote{However, the
analysis of the latest observational data shows that this conclusion
is somewhat changed, see \cite{Zhang:2007sh} for details. In this
paper, the authors derive constraints on the holographic dark energy
model from the latest observational data including the gold sample
of 182 SNIa, the CMB shift parameter given by the 3-year WMAP
observations, and the BAO measurement from the SDSS. The joint
analysis gives the fit results in 1-$\sigma$:
$c=0.91^{+0.26}_{-0.18}$ and $\Omega_{\rm m0}=0.29\pm 0.03$. That is
to say, though the possibility of $c<1$ is more favored, the
possibility of $c>1$ can not be excluded in one-sigma error range.
So, according to the new data, the evidence for the quintom feature
in the holographic dark energy model is not as strong as before.}

On the other hand, even though the current observational data
indicate that the parameter $c$ in the holographic model seems
smaller than 1, the possibility of $c\geq 1$ can not be ruled out
yet. For example, for the upper limit of the error one sigma,
$c=1.04$ in the result of the joint analysis of SNIa+CMB+LSS;
$c=1.06$ in the result of the analysis of X-ray gas data. In
particular, it must be pointed out that the choice of $c<1$, on
theoretical level, will bring some troubles. The cases of $c<1$ will
lead to dark energy behaving as a phantom eventually, which violates
the weak energy condition of general relativity,\footnote{It is
remarkable that the phantom behavior of $w<-1$ also violates the
null energy condition, which has been significant subject of
investigations, see \cite{Buniy:2005vh} for example. In
\cite{Buniy:2005vh}, the authors show that violation of the null
energy condition implies instability in a broad class of models,
which indicates for dark energy that $w$ is unlikely to be less than
$-1$.} and the Gibbons-Hawking entropy will thus decrease since the
event horizon shrinks, which violates the second law of
thermodynamics as well. Besides, the quantum instability may often
be encountered in quintom models when the $w=-1$ crossing happens.
What is more, when the future event horizon as the IR cut-off
becomes shorter than the UV cut-off within a finite time in the
future, the definition of the holographic dark energy will break
down. Consequently, from a theoretical viewpoint, the choice of
$c\geq 1$, especially of $c=1$, is more appropriate. For the favor
of $c<1$ from the currently available observational data, a possible
interpretation says that this maybe a gloss due to lack of
sufficiently precise data. Anyway, the holographic dark energy with
$c\geq 1$, especially with $c=1$, seems reasonable from theoretical
viewpoint. One the whole, since the data analysis can not rule out
the possibility of $c\geq 1$ completely, the cases of $c\geq 1$ are
worth investigating in detail. We can establish a correspondence
between the holographic dark energy with $c\geq 1$ and quintessence
scalar field, and describe holographic dark energy in this case
effectively by making use of quintessence. We refer to this case as
``holographic quintessence''.

\begin{figure}[htbp]
\begin{center}
\includegraphics[scale=0.90]{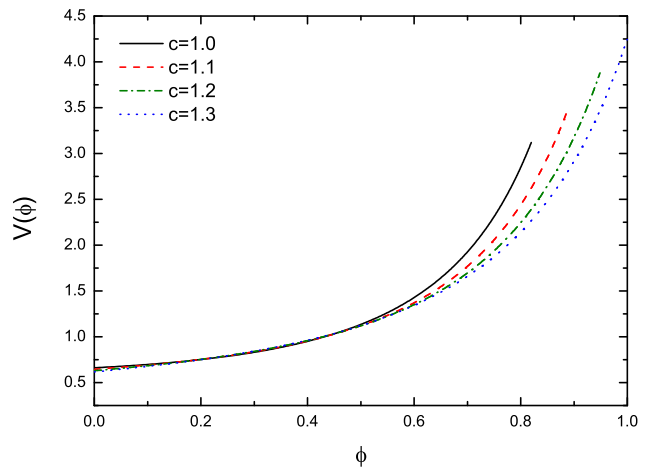}
\caption[]{\small The reconstruction of the potential for the
holographic quintessence, where $\phi$ is in unit of $M_{\rm Pl}$
and $V(\phi)$ in $\rho_{\rm co}$. We take here $\Omega_{\rm
m0}=0.3$.}\label{fig:vphi}
\end{center}
\end{figure}

\begin{figure}[htbp]
\begin{center}
\includegraphics[scale=0.90]{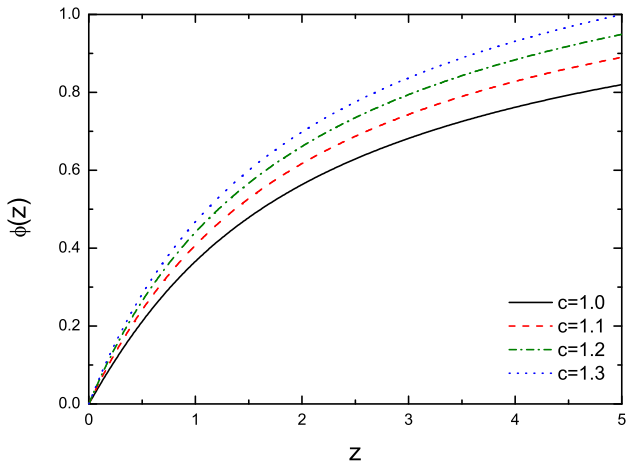}
\caption[]{\small The revolutions of the scalar-field $\phi(z)$ for
the holographic quintessence, where $\phi$ is in unit of $M_{\rm
Pl}$. We take here $\Omega_{\rm m0}=0.3$.}\label{fig:phiz}
\end{center}
\end{figure}

The quintessence potential $V(\phi)$ can be reconstructed from
supernova observational data \cite{Saini:1999ba,simplescalar}. In
addition, from some specific parametrization forms of the equation
of state $w(z)$, one can also reconstruct the quintessence potential
$V(\phi)$ \cite{Guo:2005at}. The reconstruction method can also be
generalized to scalar-tensor theories \cite{scalartensor}, $f(R)$
gravity \cite{frgrav}, a dark energy fluid with viscosity terms
\cite{viscosity}, and also the generalized ghost condensate model
\cite{Zhang:2006em}. For a reconstruction program for a very general
scalar-field Lagrangian density see \cite{general}. As discussed
above, the holographic dark energy possesses some significant
features of the quantum gravity theory. So, to some extent, we can
regard the holographic vacuum energy scenario as an underlying
theory of dark energy. The scalar-field dark energy model then can
be considered as an effective description of this holographic
theory. When $c<1$ in the holographic scenario, the quintom-like
behavior will occur, and we refer to this case as ``holographic
quintom''. The reconstruction of the scalar-field model (the
generalized ghost condensate model) according to the holographic
quintom has been investigated in detail in \cite{Zhang:2006qu}. Now
we are focussing on the reconstruction of the holographic
quintessence. We shall reconstruct the quintessence potential and
the dynamics of the scalar field in the light of the holographic
dark energy with $c\geq 1$. According to the forms of quintessence
energy density and pressure, one can easily derive the scalar
potential and kinetic energy term as
\begin{equation}
{V(\phi)\over\rho_{\rm c0}}={1\over 2}(1-w_\phi)\Omega_\phi
E^2,\label{potential}
\end{equation}
\begin{equation}
{\dot{\phi}^2\over \rho_{\rm c0}}=(1+w_\phi)\Omega_\phi
E^2,\label{kinetic}
\end{equation}
where $\rho_{\rm c0}=3M_{\rm Pl}^2H_0^2$ is today's critical density
of the universe. If we establish the correspondence between the
holographic dark energy with $c\geq 1$ and quintessence scalar
field, then $E$, $\Omega_\phi$ and $w_\phi$ are given by Eqs.
(\ref{Ez}), (\ref{deq}) and (\ref{w}). Furthermore, the derivative
of the scalar field $\phi$ with respect to the redshift $z$ can be
given
\begin{equation}
{\phi'\over M_{\rm Pl}}=\pm {\sqrt{3(1+w_\phi)\Omega_\phi}\over
1+z},\label{phiprime}
\end{equation}
where the sign is actually arbitrary since it can be changed by a
redefinition of the field, $\phi\rightarrow -\phi$. Consequently, we
can easily obtain the evolutionary form of the field
\begin{equation}
\phi(z)=\int\limits_0^z\phi'dz,\label{phi}
\end{equation}
by fixing the field amplitude at the present epoch ($z=0$) to be
zero, $\phi(0)=0$.

\begin{figure}[htbp]
\begin{center}
\includegraphics[scale=0.85]{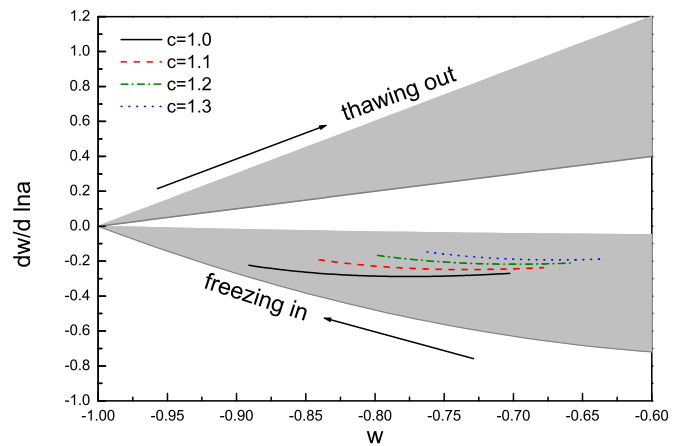}
\caption[]{\small The evolutionary trajectories of the holographic
quintessence in $w-dw/d\ln a$ phase space. The shaded regions are
occupied by thawing and freezing models respectively. The arrows
denote the evolutionary directions. The leftmost point of the
trajectories corresponds to the present; the rightmost point is at
$z=1$. As the same in previous figures, the present fractional
matter density is taken as $\Omega_{\rm m0}=0.3$. }\label{fig:phase}
\end{center}
\end{figure}

The reconstructed quintessence potential $V(\phi)$ is plotted in
figure \ref{fig:vphi}, where $\phi(z)$ is also reconstructed
according to Eqs. (\ref{phiprime}) and (\ref{phi}), also displayed
in figure \ref{fig:phiz}. Selected curves are plotted for the cases
of $c=1.0$, 1.1, 1.2 and 1.3, and the present fractional matter
density is chosen to be $\Omega_{\rm m0}=0.3$. From figures
\ref{fig:vphi} and \ref{fig:phiz}, we can see the dynamics of the
scalar field explicitly. Obviously, the scalar field $\phi$ rolls
down the potential with the kinetic energy $\dot{\phi}^2$ gradually
decreasing. The equation of state of the quintessence $w_\phi$,
accordingly, decreases gradually with the cosmic evolution, and as a
result $dw_\phi/d\ln a<0$. As suggested in \cite{Caldwell:2005tm},
quintessence models can be divided into two classes, ``thawing''
models and ``freezing'' models. Thawing models depict those scalar
fields that evolve from $w\thickapprox -1$ but grow less negative
with time as $dw/d\ln a>0$; freezing models, whereas, describe those
fields evolve from $w>-1$, $dw/d\ln a<0$ to $w\rightarrow -1$,
$dw/d\ln a\rightarrow 0$. Roughly, the holographic quintessence
should be ascribed to the freezing model. Figure \ref{fig:phase}
illustrates the freezing behavior of the holographic quintessence.
Note that it has been indicated in \cite{Caldwell:2005tm} that a
practical limit of applicability for thawing and freezing bounds
should be $w\lesssim -0.8$. Since here we only want to show the
freezing behavior of holographic quintessence in the $w-dw/\ln a$
phase space, the applicability of these regions are continued to
$w\lesssim -0.6$. As we have seen, the dynamics of the holographic
quintessence can be explored explicitly by the reconstruction.

The scalar-field models of dark energy can be viewed as low-energy
effective description of the underlying theory (e.g. quantum gravity
theory). The quintessence discussed in this paper is specified to an
ordinary scalar field minimally coupled to gravity, namely the
canonical scalar field. It is remarkable that the resulting model
with the reconstructed potential is the unique canonical
single-scalar model that can reproduce the holographic evolution
with $c\geqslant 1$ of the universe. Of course, the aforementioned
discussion can be easily generalized to other non-canonical scalar
fields, such as $K$-essence and tachyon. Moreover, it should be
noted that the holographic evolution with $c<1$ can also be
reproduced by a non-canonical scalar-field (generalized ghost
condensate), see \cite{Zhang:2006qu} for details.

In conclusion, we suggest in this paper a correspondence between the
holographic dark energy scenario and the quintessence scalar-field
model. We adopt the viewpoint of that the scalar field models of
dark energy are effective theories of an underlying theory of dark
energy. The underlying theory, though has not been achieved
presently, is presumed to possess some features of a quantum gravity
theory, which can be explored speculatively by taking into account
the holographic principle of quantum gravity theory. Consequently,
the vacuum energy acquires the dynamical property when imposing the
holographic principle. Though the currently available observational
data imply that the holographic dark energy more likely resembles a
quintom, i.e. $w$ crosses $-1$, the data analysis does not rule out
the possibility of $w>-1$ yet. Moreover, the model of $w>-1$ can
avoid some troubles the model of $w=-1$ crossing encounters. If we
regard the scalar-field model (such as quintessence) as an effective
description of such a theory (holographic vacuum energy), we should
be capable of using the scalar-field model to mimic the evolving
behavior of the dynamical vacuum energy and reconstructing this
scalar-field model according to the evolutionary behavior of
holographic dark energy. We show that the holographic dark energy
with $c\geq 1$ can be described totally by the quintessence in a
certain way. A correspondence between the holographic dark energy
and quintessence has been established, and the potential of the
holographic quintessence and the dynamics of the field have been
reconstructed.


\section*{Acknowledgements}

The author would like to thank Miao Li and Feng-Quan Wu for useful
discussions. He is also grateful to Shinji Tsujikawa and Alexander
Vikman for helpful correspondences. This work was supported in part
by the Natural Science Foundation of China.



\end{document}